\documentclass[preprint,12pt]{elsarticle}

\usepackage{amssymb}

\journal{Physica C}

\begin{document}

\begin{frontmatter}



\title{Upper critical field in electron-doped cuprate superconductor Nd$_{2-x}$Ce$_x$CuO$_{4+\delta}$: two-gap model }


\author{T.B.Charikova$^{\rm a}$\footnote{charikova@imp.uran.ru}, N.G.Shelushinina$^{\rm a}$, G.I.Harus$^{\rm a}$, D.S.Petukhov$^{\rm a}$,  V.N.Neverov$^{\rm a}$,
A.A.Ivanov$^{\rm b}$}

\address{$^{\rm a}$Institute of Metal Physics RAS, Ekaterinburg, Russia,

$^{\rm b}$Moscow Engineering Physics Institute, Moscow, Russia}

\begin{abstract}
We present resistivity measurements of the upper critical field (H$_{c2}$) phase diagram as a function  of temperature (T) for Nd$_{1.85}$Ce$_{0.15}$CuO$_{4+\delta}$/SrTiO$_3$ single crystal films with  different degree of disorder ($\delta$) in magnetic fields up to  90 kOe  at temperatures down to 0.4 K. The data are well described by a two-band/two-gap model for a superconductor in the dirty limit.

\end{abstract}


\end{frontmatter}


\section{INTRODUCTION}
The theory of the two-band superconductivity has been
developed more than 50 years ago: Suhl, Matthias, and Walker [1]
predicted the existence of {\it multigap} superconductivity, in
which a distinction of the pairing interaction in different
bands leads to different order parameters and to an enhancement of the
critical temperature.

A discovery of MgB$_2$ has renewed interest in superconductors
with multicomponent order parameters. MgB$_2$ seems to be the
first superconductor for which a two-gap model offers
a simple explanation of many anomalous experimental findings.
The two-band model for superconductivity in MgB$_2$ was first
suggested in [2,3]. On the basis of first-principles calculations
of the electronic structure and the electron-phonon
interaction, it was argued that superconductivity in this
compound resides in two groups of bands: the group of
two strongly superconducting $\sigma$-bands and the group of
two weakly superconducting $\pi$-bands.

A theory for the upper critical field H$_{c2}$ of two-gap
superconductors in the dirty limit is developed by Gurevich [4]
on the ground of a multiband BCS model [1].
The equations for H$_{c2}$ for two-band superconductors in the dirty limit are derived with the account of both intraband and interband scattering by nonmagnetic impurities. It is shown that the shape of the H$_{c2}$(T) curve essentially depends on the ratio of the intraband electron diffusivities D$_1$ and D$_2$, and can be very different from the one-gap dirty-limit theory. Taking MgB$_2$ as an example, the results of this work are used to describe  the observed abnormal temperature dependences of H$_{c2}$ of this compound.

The discovery of oxypnictides as a new class of superconductors [5] has
regenerated interest in high temperature superconductivity.
 It is shown by Mazin et al. [6] that the pairing mechanism in
iron arsenide pnictides is consistent with the multi-band
theory: the so-called ``extended  s$_{\pm}$ -wave'' model with a
sign reversal of the order parameter between different sheets of Fermi surface.
The relatively high upper critical fields (H$_{c2}$) with atypical temperature
dependence in these materials immediately attracted much
interest. In particular, the  anomalous  upward curvature observed in the temperature
dependence of H$_{c2}$ in these materials made the dirty two-band
model [4] used to describe MgB$_2$ a natural and
useful choice [7-11]. The model of Gurevich [4] is widely exploited for describing of a
variety of the abnormal H$_{c2}$ temperature dependences, both
for $H \parallel c$ and  $H \perp c$, in NdFeAsO$_{0.7}$F$_{0.3}$ single crystals [7], in polycrystalline LaFeAsO$_{0.89}$F$_{0.11}$ samples [8], in cobalt-doped SrFe$_2$As$_2$ epitaxial films [9], in Co-doped BaFe$_2$As$_2$ single crystals
[10] and in Sr$_{1-x}$Eu$_x$(Fe$_{0.89}$Co$_{0.11}$)$_2$As$_2$ single crystals
(x = 0.20 and 0.46) [11].

An anomalous upward curvature of H$_{c2}$(T) dependence has been reported in the earliest investigations of electron-doped  superconducting Nd$_{2-x}$Ce$_x$CuO$_{4+\delta}$ crystals [12, 13]. Such a non-BCS behavior  of H$_{c2}$(T) was observed later both in Nd$_{2-x}$Ce$_x$CuO$_{4+\delta}$ single crystals [14] and thin films [15, 16].

Gantmacher et al. [14] have argued that the observed concave form of  H$_{c2}$(T)   in their bulk 
Nd$_{1.82}$Ce$_{0.18}$CuO$_{4+\delta}$   crystals may be an indication of unconventional boson-type superconductivity. In the work of Golnik and Naito [15] it is shown that the normal state resistivity, Hall coefficient  and magnetoresistivity of Nd$_{2-x}$Ce$_x$CuO$_{4+\delta}$  films can be described quantitatively within a simple two-carrier model if the existence of an electronlike and a holelike band is assumed. The physical origin of the two conduction bands at that time remained an open question. In the superconducting regime the positive curvature of the resistive critical field was attributed by them to the quasi-2D character of the material and the resulting fluctuation effects. 

There are much recent activities in investigation of possible electrons and
holes coexistence in the normal state of cuprate superconductors [17]. Two kinds of carriers in electron-doped cuprates with different lanthanide cations seem to arise from the electronic structure near the Fermi surface (FS) of the CuO$_2$ planes.
 Angle resolved photoemission spectroscopy (ARPES) has revealed a small electron-like FS pocket in the underdoped region, and a simultaneous presence of
both electron- and hole-like pockets near optimal doping in Nd$_{2-x}$Ce$_x$CuO$_{4+\delta}$  systems [18,19]. 
The conclusions of the ARPES measurements and first-principle
 calculations of the electronic
structure on the electron-doped high-Tc superconductors Ln$_{1.85}$Ce$_{0.15}$CuO$_4$
(Ln = Nd, Sm and Eu) performed by Ikeda et al. [20], are in accordance
with these results.
A spin density wave (SDW) model [21-23] was proposed which gives qualitative explanation to ARPES observations. In this model, SDW ordering
would induce FS reconstruction that results in an evolution from an electron
pocket to the coexistence of electron-like and hole-like pockets and then into
a single hole-like FS with increasing of doping.

In our previous work [24]
we have analyzed the Ce -doping dependence of the normal state Hall
coeffient in optimally reduced Nd$_{2-x}$Ce$_x$CuO$_{4+\delta}$/SrTiO$_3$  single crystal films and, on
the grounds of this analysis, we have recruited a two-carrier model for describing
of the mixed state Hall coeffcient. We have adopted a semi-phenomenological description of a mixed state Hall effect by flux-flow model of Bardeen and Stephen modified by coexistence of electrons and holes.
In this work we present results on the upper critical field as a function of temperature determined by a resistive method in Nd$_{2-x}$Ce$_x$CuO$_{4+\delta}$/SrTiO$_3$ single crystal films with x= 0.15 and various degree of disorder ($\delta$). Our goal is to describe the observed abnormal H$_{c2}$ (T) dependences on the ground of two-gap model of Gurevich [4]. Such an approach may occur
actual in the light of much recent efforts on experimental (ARPES) and
theoretical (SDW-model) investigations of the electronic structure near Fermi
surface in the CuO$_2$ plane of electron-doped cuprate superconductors.

\section{SAMPLES AND EQUIPMENT}

The series of Nd$_{2-x}$Ce$_x$CuO$_{4+\delta}$/SrTiO$_3$ epitaxial films ($x$ =   0.15) with standard (001) orientation were synthesized by pulsed laser evaporation [16]. The original target (the sintered ceramic tablet of Nd$_{2-x}$Ce$_x$CuO$_{4+\delta}$ of the given composition) was evaporated by a focused laser beam and the evaporated target material was deposited on a heated  single-crystal substrate. The substrate material was SrTiO$_3$ with (100) orientation and dimensions of 5$\times$10$\times$1.5 mm. The substrate temperature was 800$^0$\,C, the pressure during the deposition was 1.067 mbar, the residual gas was nitrous oxide (N$_2$O). Then the films were subjected to heat treatment (annealing) under various conditions to obtain samples with various oxygen content.  As a result, three types of samples with $x$=0.15 were obtained: as-grown samples, optimally reduced samples (optimally annealed in a vacuum at T = 780$^0$\,C for t = 60 min; p = 1.33$\cdot$10$^{-2}$\,mbar) and non optimally reduced samples (annealed in a vacuum T = 780$^0$\,C for t = 40 min; p = 1.33$\cdot$10$^{-2}$\,mbar). Temperature dependence of resistivity were measured at the magnetic field up to 90 kOe using the Quantum Design PPMS in the temperature range $T =$ (1.8 $\div$ 300) K and the Oxford Instruments superconducting magnet up to  120 kOe in the temperature range $T =$ (0.4 $\div$ 4.2) K.

\section{EXPERIMENTAL RESULTS AND DISCUSSION}

The in-plane longitudinal resistivity $\rho_{xx}$ is measured as a function of temperature T for different applied magnetic fields H or as a function of H at fixed temperatures in single crystal films of electron-doped superconductor
Nd$_{2-x}$Ce$_x$CuO$_{4+\delta}$ in magnetic field perpendicular to the ab-plane up to 90 kOe at T=(0.4 $\div$ 300)K. Fig.1a shows the $\rho$(T) dependences for the optimally reduced Nd$_{1.85}$Ce$_{0.15}$CuO$_4$ film. It is seen
that  an increase of the external magnetic field results in a
shift of the resistive transition to lower temperatures without
appreciable broadening. Thus the values of the upper critical field H$_{c2}$ can be deduced from $\rho$(T)) curves for fixed magnetic fields. For the definition of H$_{c2}$ we use  a value that is the intersection of  tangents to the resistive transition region and  to the normal-state resistivity region  (see Fig.1b).

\begin{figure}
\includegraphics{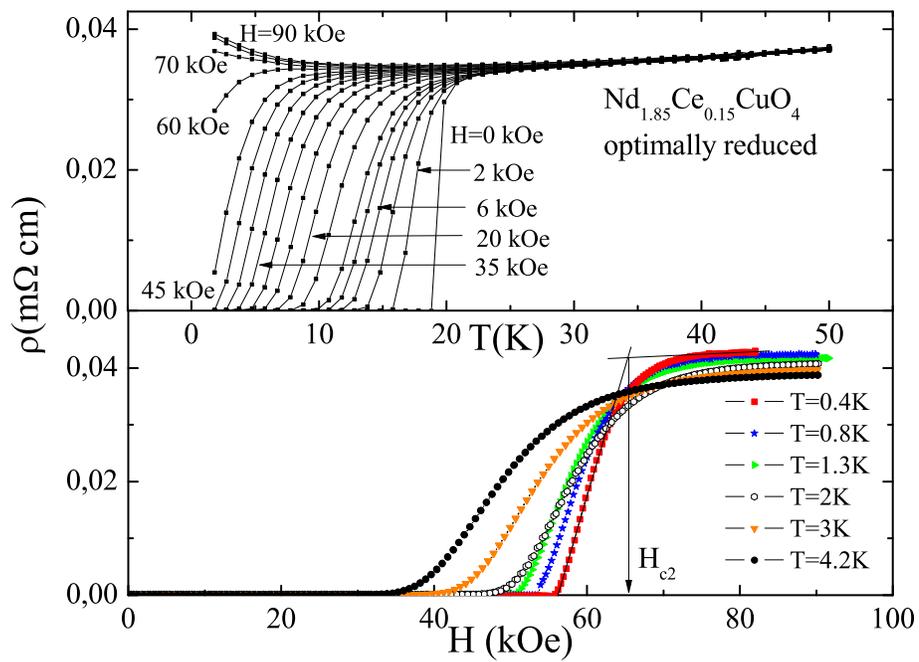}
\caption{\label{fig:wide} The resistive transitions of the Nd$_{1.85}$Ce$_{0.15}$CuO$_4$/SrTiO$_3$ film with optimal annealing: (a) as function of temperature at various magnetic fields; (b) as a function of magnetic field at various temperatures.}
\end{figure}

In Fig.2 we have plotted H$_{c2}$(T) dependences derived from measured $\rho$(T) and $\rho$(B)
transitions for all our Nd$_{1.85}$Ce$_{0.15}$CuO$_{4+\delta}$  samples. In both optimally and nonoptimally reduced  films we observe a positive curvature of H$_{c2}$(T)  for temperatures from T$_c$ down to about T$_c$/2
and then obvious tendency to saturation at low temperatures.

\begin{figure}
\includegraphics{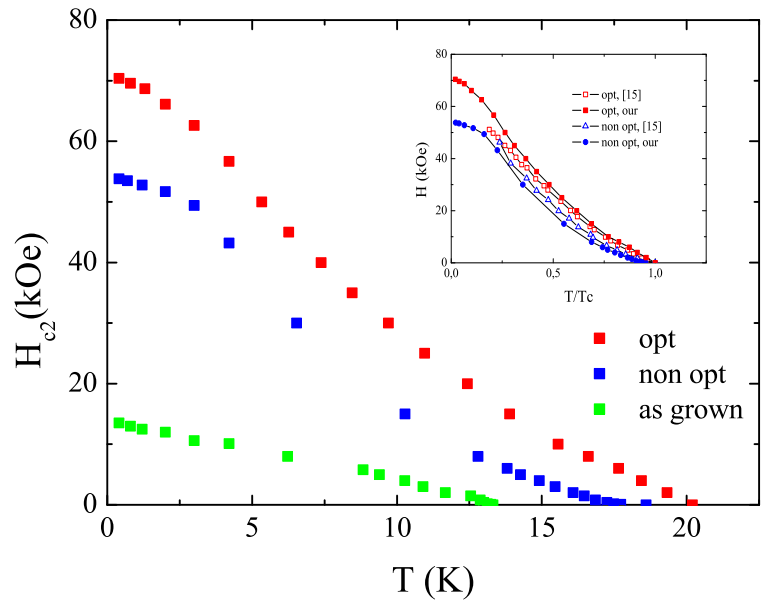}
\caption{\label{fig:wide} Temperature dependences of the upper critical field H$_{c2}$ defined as a value of the crossing of two tangents: first - to the resistive transition region; second - to the normal resistivity region. Inset: H$_{c2}$(T) dependences for optimally and nonoptimally reduced samples of  Golnik and Naito [15] - open labels; our experimental data of H$_{c2}$(T) at T$<$4.2K - solid labels.}
\end{figure}

On the insert of Fig.2 H$_{c2}$(T) dependences for optimally and nonoptimally reduced samples of  Golnik and Naito [15] are presented for comparison. It is seen that their and our dependences have 
similar behavior at T $>$ 4.2K and that our data rather well complement the data of [15] at
T $<$ 4.2 K. Measurements of the resistive critical field of Nd$_{1.84}$Ce$_{0.16}$CuO$_{4+\delta}$  single crystals
have been reported for temperatures below 4.2 K with similar results (see [12,13]).

As revealed by ARPES [18,19] the two types of carriers with opposite charges in the normal state of electron-doped cuprate comes from two parts of FS that is
associated with a reconstruction of conduction band due to SDW ordering which splits the
band into upper and lower branches [25, 26].  
The inset of Fig. 3a shows the FSs for a typical optimally-doped 
sample [22]. The SDW correlation splits the continuum FS
into two pieces of sheet: electron-like(1) and hole-like (2) pockets. 

In the SC state  the reconstruction of Fermi surface  leads naturally to a two-gap model.
The situation with piecewise FS (with two types of FS parts) in
superconducting state of electron doped cuprates is properly considered, for
example, by Liu and Wu [22]. The model suggests two SC gaps
 with different amplitude: $\Delta$$_1$  for electron-like band and $\Delta$$_2$   for hole-
like one. The two-gap model gives a unified explanation for the  experimental data on ARPES [19] and Raman scattering [27] in the electron-doped
cuprates with x $<$ 0.17.

On these grounds we explore theoretical considerations about temperature dependence of the upper critical field in two-gap superconductor [4] for an interpretation of our experimental data.

We regard  Nd$_{2-x}$Ce$_x$CuO$_{4+\delta}$  single crystal with x= 0.15 as a superconductor
with two sheets (pockets) 1 and 2 of the Fermi surface on which
the superconducting gaps take the values $\Delta$$_1$  and $\Delta$$_2$, respectively
(indices 1 and 2 correspond to electron and hole pockets) and 
the intraband  diffusivities electron D$_1$ and hole D$_2$.

The equation for H$_{c2}$ for two-band/ two-gap superconductor in the dirty limit are derived in [4] with the account of both intraband and interband scattering by nonmagnetic impurities and paramagnetic effects. For a simpler case of negligible interband scattering and paramagnetic effects
the equation for H$_{c2}$   has been presented as [4b]:

\begin{eqnarray}
\lefteqn{ ln(t) = -[U(h)+U(\eta h)+\lambda_{0}/w]/2 +} \nonumber\\
&&+[(U(h)-U(\eta h)-\lambda_{-}/w)^2/4+\lambda_{12}\lambda_{21}/w^2]^{1/2}.
\end{eqnarray}

Here $t$ = T/T$_c$, $h$=H$_{c2}$D$_1$/2$\Phi_0$T,  $\Phi_0$ = $\pi \hbar$/e -  is a flux quantum, $\eta$ = D$_2$/D$_1$, 

\begin{equation}
U(x)= \psi(1/2+x)- \psi(1/2),
\end{equation}

where  $\psi$(x) is a digamma function and $\lambda_{\pm}$ = $\lambda_{11}$ $\pm$ $\lambda_{22}$, $w$= $\lambda_{11}\lambda_{22}$ - $\lambda_{12}\lambda_{21}$, $\lambda_0$ =($\lambda_{-}^{2}$ + 4$\lambda_{12}\lambda_{21}$)$^{1/2}$.

Eq.(1)contains the matrix of the BCS coupling constants: the diagonal terms $\lambda_{11}$ and $\lambda_{22}$ describe intraband pairing (band 1 have the highest coupling constant $\lambda_{11}$) and  $\lambda_{12}$ and $\lambda_{21}$ describe interband coupling.

For equal diffusivities, $\eta$ = 1, Eq. (1) simplifies to the one-gap de-Gennes-Maki equation of Werthamer - Helfand - Hohenberg  (WHH) theory (see [28] and references therein):

\begin{equation}
ln(t)+U(h)=0.
\end{equation}

As was to be expected for $\lambda_{12}$=0 equation (1) breaks up on two independent WHH ones. For the first type of carriers with D=D$_1$ we have:

\begin{equation}
ln(t^{(1)})+U(h)=0,
\end{equation}

where t$^{(1)}$ =T/T$_c^{(1)}$, T$_c^{(1)}$ $\sim$ exp(-1/$\lambda_{11}$). For the second type with D=D$_2$ we obtain from (1):

\begin{eqnarray}
ln(t^{(1)})+U(\eta h)= \lambda_0/w,\nonumber\\
or \nonumber\\
ln(t^{(2)})+U(\eta h)=0,
\end{eqnarray}

where t$^{(2)}$ = T/T$_c^{(2)}$,T$_c^{(2)}$$\sim$ exp(-1/$\lambda_{22}$) and expressions $\lambda_0$/$w$=($\lambda_{11}$ - $\lambda_{22}$)/($\lambda_{11}\lambda_{22}$) and
T$_c^{(2)}$/T$_c^{(1)}$ = exp(-($\lambda_{11}$ - $\lambda_{22}$)/($\lambda_{11}\lambda_{22}$)), reliable at $\lambda_{12}$ = 0, are used.

On Figs 3, 4 we show the fit of the two-band theoretical equation (1) to H$_{c2}$(T) curves with fitting parameters listed in Table I for optimally (Fig. 3a)  and non optimally (Fig. 4a) reduced Nd$_{1.85}$Ce$_{0.15}$CuO$_{4+\delta}$  samples. We find that for both samples the curves of
H$_{c2}$(T) can be rather well described by the two-band model for the diffusivity ratio 
$\eta$ = D$_2$/D$_1$$ <$ 1 and for nearly negligible interband BCS coupling constants: $\lambda_{12}$ $\ll$  $\lambda_{11}$, $\lambda_{22}$.
It follows from the previous transport measurements on two investigated samples [24] that carriers with the higher diffusivity D$_1$ are electrons and ones with the lower diffusivity D$_2$ are holes.

\begin{figure}
\includegraphics{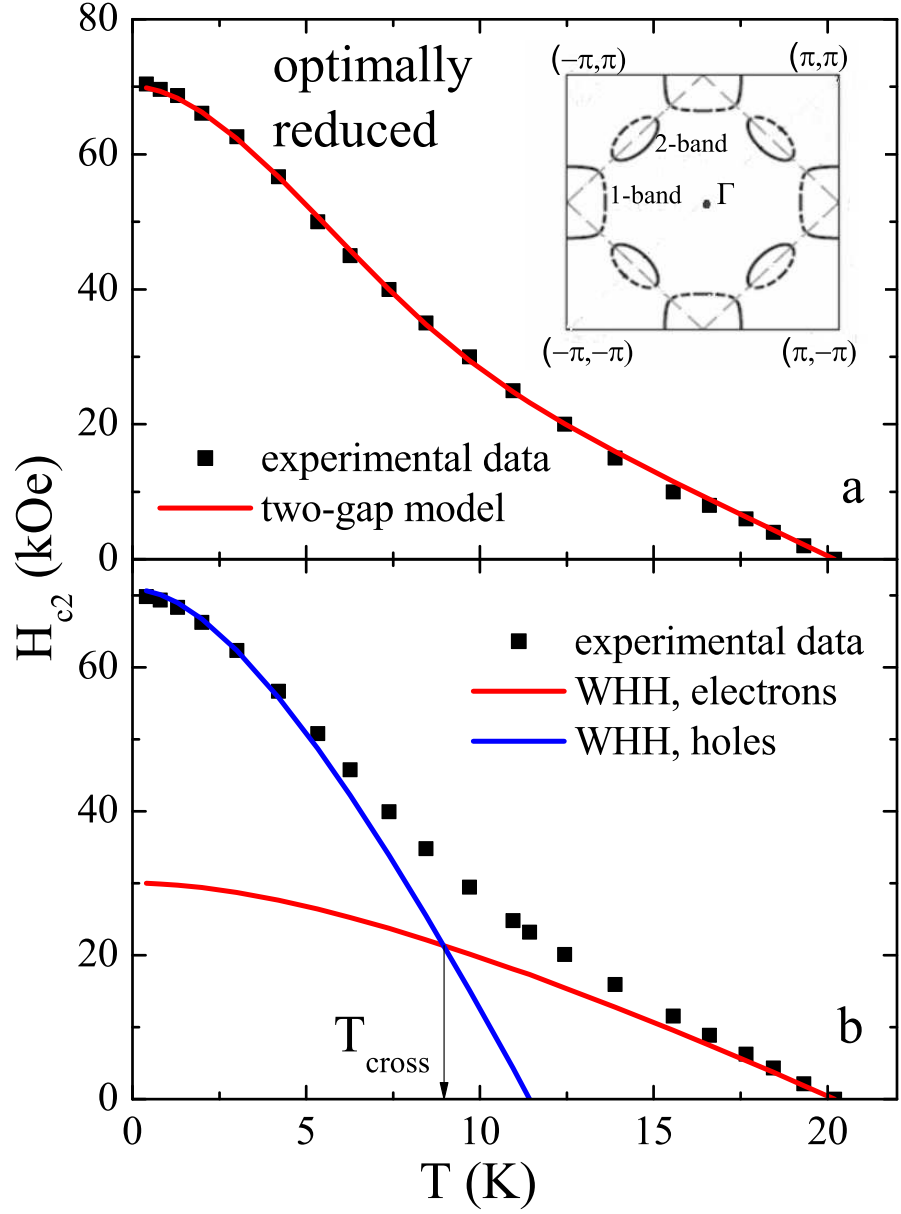}
\caption{\label{fig:wide} Upper critical field versus temperature for optimally reduced Nd$_{1.85}$Ce$_{0.15}$CuO$_{4}$  sample. The filled symbols correspond to the experimental data.
(a) Solid line corresponds to H$_{c2}$(T) calculated from the two-gap Eq.(1) for the  fitting parameters listed in Table 1. (b) Red and blue curves are calculated according to the one-gap Eqs (4) and (5), respectively. The  fitting parameters are listed in Table 2.}
\end{figure}

\begin{figure}
\includegraphics{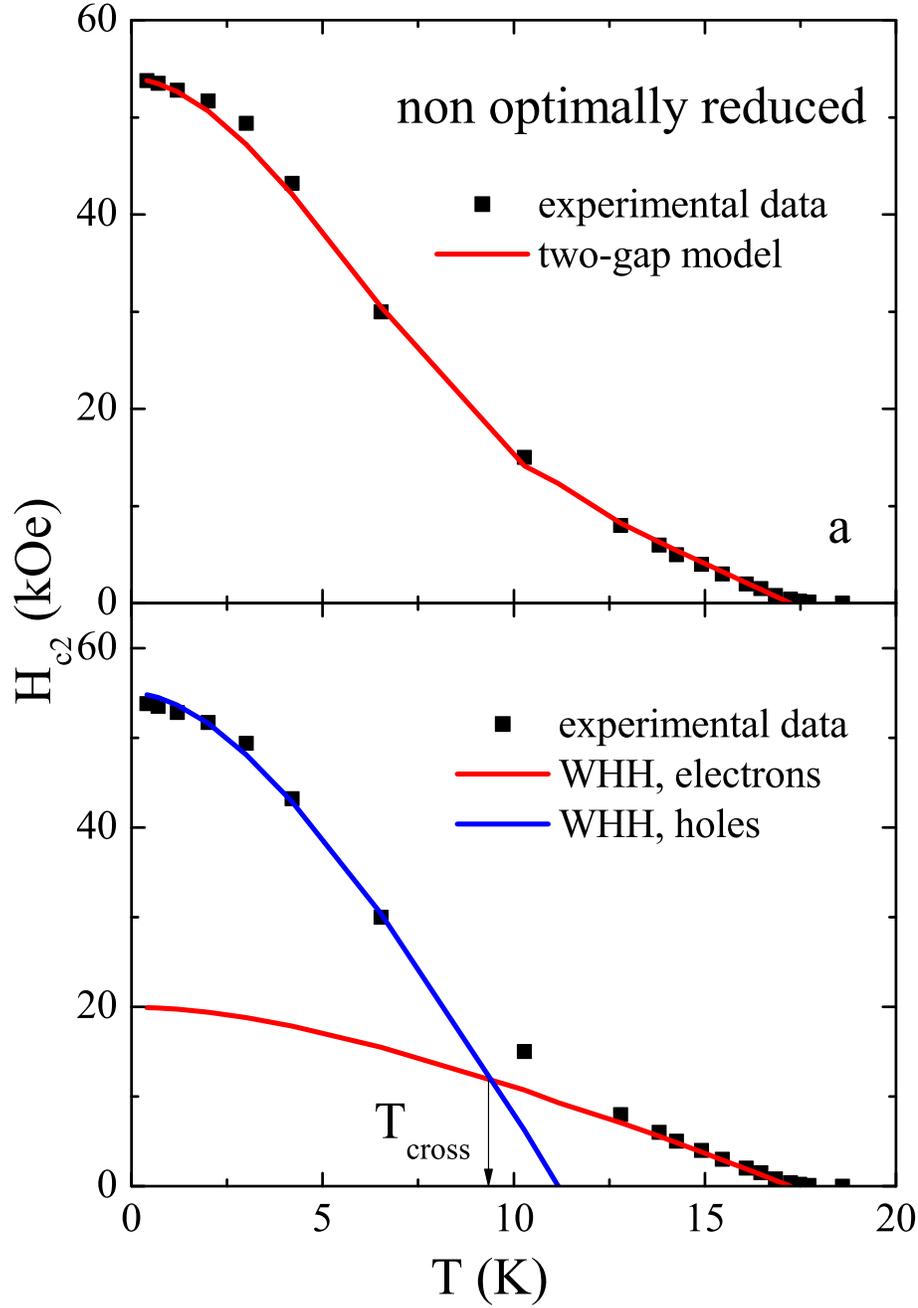}
\caption{\label{fig:wide} Upper critical field versus temperature for non optimally reduced Nd$_{1.85}$Ce$_{0.15}$CuO$_{4+\delta}$  sample. The filled symbols correspond to the experimental data.
(a) Solid line corresponds to H$_{c2}$(T) calculated from the two-gap Eq.(1) for the  fitting parameters listed in Table 1. (b) Red and blue curves are calculated according to the one-gap Eqs (4) and (5), respectively. The  fitting parameters are listed in Table 2.}
\end{figure}

     Encouraging by the strong inequality between interband and intraband coupling constants
we try to describe H$_{c2}$(T) dependences on the ground of independent equations (4) and (5)
for $\lambda_{12}$ = 0. The fitting for optimally (Fig. 3b) and non optimally (Fig. 4b) reduced samples shows that the observed dependences may be considered as a superposition of WHH-like curves 
for two types of carriers (electrons and holes) with different critical fields  H$_{c2}^{(1)}$ or H$_{c2}^{(2)}$  and  with different critical temperatures T$_{c}^{(1)}$ or T$_{c}^{(2)}$. The fitting parameters are presented in Table 2. As it is seen from Figs. 3b and 4b it is necessary to take into account the finite value of $\lambda_{12}$ only near a crossover of  the two WHH curves at T = T$_{cross}$.

     In accordance with analysis of Gurevich [4] in a case of different diffusivities the limiting value of H$_{c2}$(0) is determined by the minimum diffusivity (i.e. by holes):

\begin{equation}
H_{c2}(0)=\Phi_{0}T_{c}^{(2)}/2\gamma D_2,
\end{equation}

where $\gamma \cong$ 1.78. On the other hand the expressions for H$_{c2}$(T) near T$_c$ is determined by the band with the
highest coupling constant $\lambda_{11}$ (i.e. by electrons):

\begin{equation}
H_{c2}(T)=4\Phi_{0}(T_{c}-T)/\pi^2 D_1.
\end{equation}

    The experimentally observed ("real") values of H$_{c2}$ and T$_c$ should be the highest ones: for
critical fields these are H$_{c2}^{(2)}$  at T $<$ T$_{cross}$  and  H$_{c2}^{(1)}$ at T $>$ T$_{cross}$, for critical temperature T$_c$ = T$_c^{(1)}$ (see Figs 3b, 4b). The lower ("virtual" in terms of the article [4b]) values of H$_{c2}$ come to the light, for example, in the magnetic field dependences of Hall coefficient, R$_H$(B), in a mixed state of superconductor (see [24]).

   On Fig.5 we present the data for H$_{c2}$(T)  dependences in as grown Nd$_{1.85}$Ce$_{0.15}$CuO$_{4+\delta}$ sample.
In contrast to the other two samples we observe here a convex form of H$_{c2}$(T) curve at all temperature interval.  The data are remarkably well described by the WHH-like equation in
accordance with a result of transport measurements [24] that electrons and holes give nearly
equal contributions to the conductivity in this sample, i.e. D$_1$ $\cong$ D$_2$. It is known [4] that for equal diffusivities of the two bands, i.e., $\eta$ = D$_2$/D$_1$ = 1, the parametric equation (1) reduces to the one-gap formula in the WHH theory.

\begin{figure}
\includegraphics{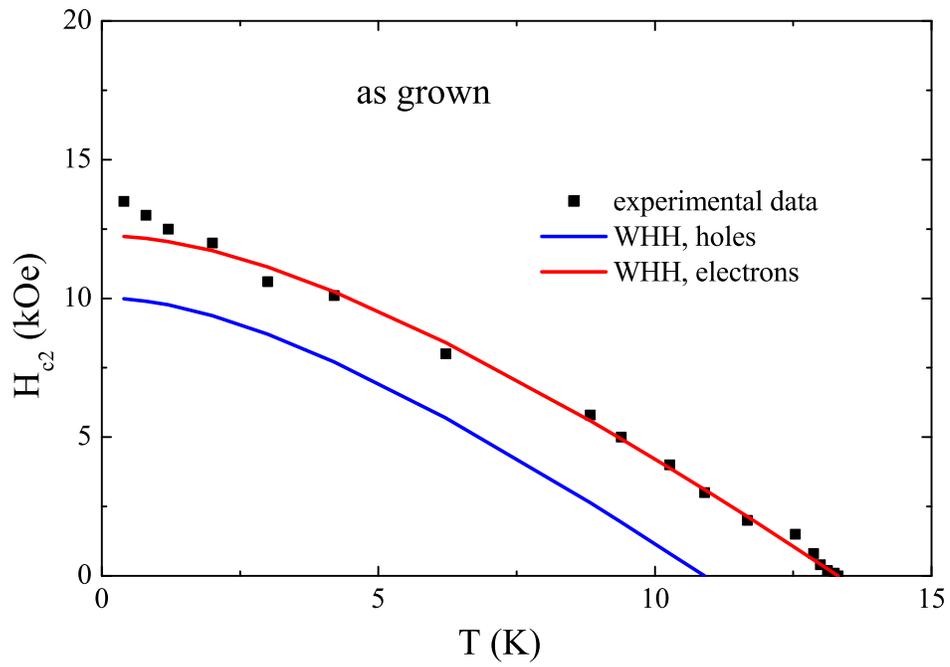}
\caption{\label{fig:wide} Upper critical field versus temperature for as grown Nd$_{1.85}$Ce$_{0.15}$CuO$_{4+\delta}$  sample. The filled symbols correspond to the experimental data.
Red and blue curves are calculated according to the WHH formulas (4) and (5), respectively. The  fitting parameters are listed in Table 2.}

\end{figure}

    From an analysis of R$_H$(B) dependences in a mixed state [24] we have obtained that in as grown  sample the critical field for electrons is higher than that for holes: H$_{c2}^{(1)}$ $>$ H$_{c2}^{(2)}$, and both real (H$_{c2}^{(1)}$) and virtual(H$_{c2}^{(2)}$) fields  have been determined (see Table 2). As for equal diffusivities one has T$_c^{(2)}$/T$_c^{(1)}$ = H$_{c2}^{(2)}$/H$_{c2}^{(1)}$ , we can constructed not only real WHH curve for electrons but also virtual one for holes (see Fig.5).

It should be noted that the existence of two different spatial and magnetic field scales, consistent with the experimental data on Abrikosov vortex structure in MgB$_2$, has been demonstrated by Koshelev
and Golubov [29]. 

They provided a quantitative model for the vortex structure in a two-band superconductor with weak interband impurity scattering and strong intraband scattering rates.
It was shown that in a case of weak interband scattering the two typical sizes of the isolated vortex in different bands appear and, as a consequence, two magnetic field scales exist which can be accessible experimentally.

\section{CONCLUSIONS}

In conclusion, we report the data for the upper critical field as a function of temperature for Nd$_{1.85}$Ce$_{0.15}$CuO$_{4+\delta}$ single crystal films with  different degree of disorder ($\delta$) determined by the resistivity measurements. Two types of H$_{c2}$(T) dependences are observed: the abnormal ones with a positive curvature of H$_{c2}$(T)  for temperatures from T$_c$ down to about T$_c$/2 in both optimally and non optimally reduced  films and convetional WHH curve in as grown film.
We find that the H$_{c2}$(T) dependences can be consistently explained by the two-band/two-gap model of a dirty superconductor for the diffusivity ratio $\eta$ = D$_2$/D$_1$ $<$ 1  in the first two films and for equal diffusivities, $\eta$ = D$_2$/D$_1$ = 1, in the third one.

Thus, we have demonstrated that the two-band/two-gap model captures the essence of the behavior of H$_{c2}$(T) for investigated systems. This result is in accordance with experimental evidences and theoretical considerations for a coexistence of electrons and holes in a normal
state of optimally doped Nd$_{2-x}$Ce$_x$CuO$_{4+\delta}$ system.

This work was done within RAS Program (project N 12-P-2-1018) with partial support of RFBR (grant N 12-02-00202).



REFERENCES



\begin{thebibliography}{29}

\bibitem{1} H. Suhl, B.T. Matthias, and L.R. Walker, Phys. Rev. Lett.{\bf 3}, 552 (1959).
\bibitem{2} A. Y. Liu, I. I. Mazin, and J. Kortus, Phys. Rev. Lett.{\bf 87}, 87005 (2001).
\bibitem{3} S. V. Shulga et. al., cond-mat/0103154.
\bibitem {4} (a) A. Gurevich, Phys. Rev. B {\bf 67} 184515 (2003); (b) A. Gurevich, Physica C {\bf 456}, 160 (2007). 
\bibitem {5} Y. Kamihara, T. Watanabe, M. Hirano, and H. Hosono: J. Am. Chem.Soc.{\bf 130} 3296 (2008).
\bibitem {6} I. I. Mazin, D. J. Singh, M. D. Johannes, and M. H. Du: Phys. Rev.Lett. {\bf 101} 057003 (2008).
\bibitem {7} J. Jaroszynski, F. Hunte, L. Balicas, Youn-jung Jo, I. Raicevich, A. Gurevich, and D. C. Larbalestier, F. F. Balakirev, L. Fang, P. Cheng, Y. Jia, and H. H. Wen, Phys. Rev. B {\bf 78}, 174523 (2008).
\bibitem {8} F. Hunte, J. Jaroszynski, A. Gurevich, D. C. Larbalestier, R. Jin, A. S. Sefat, M. A. McGuire, B. C. Sales, D. K. Christen,  D. Mandrus, Nature (London) {\bf 453}, 903 (2008).
\bibitem {9} S. A. Baily, Y. Kohama,H. Hiramatsu, B. Maiorov, F. F. Balakirev, M. Hirano, and H. Hosono, Phys. Rev. Lett. {\bf 102}, 117004 (2009). 
\bibitem {10} M. Kano, Y. Kohama, D. Graf, F. Balakirev, A. S. Sefat, M. A. Mcguire, B. C. Sales, D.Mandrus, and S. W. Tozer, J. Phys. Soc. Jpn. {\bf 78}, 084719 (2009).
\bibitem {11} Rongwei Hu, Eun Deok Mun, M. M. Altarawneh, C. H. Mielke, V. S. Zapf, S. L. Bud'ko, and P. C. Canfield, Phys. Rev. B {\bf 85}, 064511 (2012).
\bibitem {12} Y. Hidaka and M. Suzuki, Nature (London) {\bf 338}, 635 (1989).
\bibitem {13} Y. Dalichaouch, B. W. Lee, C. L. Seaman, J. T. Markert, and M.
B. Maple, Phys. Rev. Lett. {\bf 64}, 599 (1990).
\bibitem {14} V.F.Gantmacher, G.A. Emel'chenko, I.G. Naumenko, G.E. Tsydynzhapov, JETP Letters
{\bf 72}, 21 (2000).
\bibitem {15} F. Gollnik, M. Naito, Phys. Rev. B {\bf 58}, 11734 (1998).
\bibitem {16} T. B. Charikova, N. G. Shelushinina, G. I. Kharus, and A.A.Ivanov, JETP Letters {\bf88}, 123 (2008). 
\bibitem {17} N.Luo, arXiv:cond-mat/0003074v2.
\bibitem {18} N. P. Armitage, F. Ronning, D. H. Lu, C. Kim, A. Damascelli, K. M.
Shen, D. L. Feng, H. Eisaki, and Z.-X. Shen, P. K. Mang, N. Kaneko,
and M. Greven, Y. Onose, Y. Taguchi, and Y. Tokura, Phys.Rev.Lett.
{\bf 88}, 257001 (2002); N.P.Armitage, P. Fournier, and R. L. Greene, Rev.Mod.Phys. {\bf 82}, 2421
(2010).
\bibitem {19} H. Matsui, K. Terashima, T. Sato, T. Takahashi, S.-C. Wang, H.-B.
Yang, H. Ding, T. Uefuji, and K. Yamada, Phys.Rev.Lett. {\bf 94}, 047005
(2005); H. Matsui, K. Terashima, T. Sato, T. Takahashi, M. Fujita, and K.
Yamada, Phys.Rev.Lett. {\bf 95}, 017003 (2005).
\bibitem {20} M. Ikeda, T. Yoshida, A. Fujimori, M. Kubota, K. Ono, K. Unozawa ,
T. Sasagawa, H. Takagi, J.Supercond.Nov.Magn. {\bf 20}, 563 (2007); M. Ikeda, T. Yoshida, A. Fujimori, M. Kubota, K. Ono, Hena Das, T.Saha-Dasgupta, K. Unozawa, Y. Kaga, T. Sasagawa, and H. Takagi, Phys.Rev.B {\bf 80}, 014510 (2009).
\bibitem {21} J. Lin and A. J. Millis, Phys.Rev.B {\bf 72}, 214506 (2005).
\bibitem {22} C. S. Liu and W. C. Wu, Phys.Rev.B {\bf 76}, 014513 (2007), Figs.1c,d.
\bibitem {23}E. Z.Kuchinskii, M.V. Sadovskii, JETP Letters {\bf 88}, 224 (2008).
\bibitem {24} T.B. Charikova, N.G. Shelushinina, G.I. Harus, D.S. Petukhov, A.V. Korolev, V.N. Neverov, A.A. Ivanov, Physica C {\bf 483}, 113 (2012).
\bibitem {25} C. Kusko, R. S. Markiewicz, M. Lindroos, and A. Bansil, Phys. Rev. B {\bf 66}, 140513(R) (2002).
\bibitem {26} Q. Yuan, Y. Chen, T. K. Lee, and C. S. Ting, Phys. Rev. B {\bf 69},
214523  (2004).
\bibitem {27} G. Blumberg, A. Koitzsch, A. Gozar, B. S. Dennis, C. A.
Kendziora, P. Fournier, and R. L. Greene, Phys. Rev. Lett. {\bf 88},
107002  (2002).
\bibitem {28} N.R. Werthamer, E. Helfand, P.C. Hohenberg, Phys. Rev. {\bf 147}, 295
(1966).
\bibitem {29} A. E. Koshelev A. A. Golubov , Phys. Rev. Lett. {\bf 90}, 177002 (2003). An equation (11) of [29] is similar to Eq. (34) reported  in Ref. 4(a) or Eq. (25) of Ref. 4(b) (see Eq. (1) in our work).




\end{thebibliography}






new page
\begin{table}[b]
\caption{\label{tab:table1}
Parameters of the fits to the two-gap formula (1) for Nd$_{1.85}$Ce$_{0.15}$CuO$_{4+\delta}$/SrTiO$_3$ epitaxial films.}

\begin{tabular}{ccccccc}
\hline
Samples & T$_c$,  & H$_{c2}$(0),  & $\lambda_{11} \lambda_{12}\choose \lambda_{21} \lambda_{22}$ & D$_1$& D$_2$& $\eta$  \\
        & K     & kOe           & &      &      &         \\          
\hline
Optimally& 20.2 & 70.4  & ${0.43 ~ 0.044 \choose {0.044 ~ 0.35}}$  & 1.89 & 0.45  & 0.24 \\
reduced  &  &   &    &  &   &  \\
\hline      
Non optomally& 17.3 & 53.8   & ${0.54 ~ 0.038 \choose {0.038 ~ 0.44}}$  &  2.43  & 0.57  & 0.235 \\
reduced      &  &    &  &    &   &  \\
\hline
\end{tabular}
\end{table}

\begin{table}[b]
\caption{\label{tab:table2}
Parameters of the fits to the WHH-like formulas (4) and (5)for Nd$_{1.85}$Ce$_{0.15}$CuO$_{4+\delta}$/SrTiO$_3$ epitaxial films.}

\begin{tabular}{cccccccc}
\hline
Samples & T$_c^{(1)}$,  & H$_{c2}^{(1)}$(0),  & T$_c^{(2)}$& H$_{c2}^{(2)}$(0) & D$_1$& D$_2$& $\eta$  \\
        & K     & kOe           & K &   kOe   &      &      &   \\          
\hline
Optimally& 20.2 & 30.0  & 11.4  & 70.4 & 1.89  & 0.45 & 0.24 \\
reduced  &  &   &    &  &   & &  \\
\hline      
Non optimally& 17.3 & 20.0  & 11.2  &  53.8 & 2.43 & 0.57 & 0.235 \\
reduced      &  &    &  &    &   & &  \\
\hline
As grown & 13.3  & 12.2 & 10.9 &  10.0  & 3.0  & 3.0 & 1 \\
\hline
\end{tabular}
\end{table}

\end{document}